\documentclass[conference]{IEEEtran}
\IEEEoverridecommandlockouts
\usepackage{cite}
\usepackage{amsmath,amssymb,amsfonts}
\usepackage{algorithmic}
\usepackage{graphicx}
\usepackage{textcomp}
\usepackage{xcolor}
\usepackage{algorithm} 
\usepackage{hyperref}
\usepackage{xurl}
\def\BibTeX{{\rm B\kern-.05em{\sc i\kern-.025em b}\kern-.08em
    T\kern-.1667em\lower.7ex\hbox{E}\kern-.125emX}}
\begin{document}

\title{A Data-Driven Filter Bank Framework for IMU-Based Heave Motion Estimation\\

\thanks{\textbf{This work was supported by ASELSAN Inc.}}
}

\author{\IEEEauthorblockN{Aybars Tokta}
\IEEEauthorblockA{\textit{Avionic Systems Engineering Directorate} \\
\textit{ASELSAN Inc.}\\
Ankara, Türkiye \\
 atokta@aselsan.com}


}

\maketitle

\begin{abstract}
In this study, we address IMU-based heave motion estimation problem for inertial navigation systems. Unlike existing approaches, we propose a data-driven framework in which a bank of IIR filters, each associated with a specific frequency range, is optimized using a synthetically generated dataset of realistic heave–acceleration tuples. The synthetic heave signal generation pipeline starts by synthesizing random wave signals from established wave energy spectra and then processing them through heave response amplitude operators reported in the literature. The corresponding vertical acceleration measurements are obtained by double-differentiating the heave signals and corrupting them with realistic low- and high-frequency disturbances observed in real IMU recordings. A Fourier transform–based method is used to estimate the mean peak period and select the appropriate filter. Simulation results from both offline and real-time tests demonstrate that the proposed method is robust to varying sea regimes and provides accurate heave estimation, with a maximum RMSE not exceeding the larger of 5 cm or 5\% of the significant heave height.
\end{abstract}


\section{Introduction}
A vessel at sea is subjected to irregular wave excitations, which cause it to move in six degrees of freedom. Estimating a ship’s vertical position with respect to the local waterline level, known as heave, is critical for a wide range of marine applications, from active heave compensation systems used in crane operations such as offshore lifting and underwater installations, to precise aerial vehicle landings on ship decks, as well as radar motion compensation \cite{b1}, \cite{b2}. In the literature, different approaches exist for estimating a ship’s heave motion depending on the sensor type. One of these approaches is based on filtering standalone IMU measurements, where the vertical acceleration is double-integrated and passed through a high pass filter to eliminate low frequency errors present in the computed vertical acceleration due to biases, tilt errors, and other error sources \cite{b3}, \cite{b4}.  There are also Kalman filter based methods available in the literature. Richter \textit{et al}. propose an observer that relies solely on acceleration measurements and estimates the harmonic components of the heave motion, where each harmonic is represented by three states in the dynamic model \cite{b5}. In \cite{b6}, GPS RTK measurements are fused with vertical accelerometer measurements. More recently, AI-based models have been developed to estimate ship motions including heave \cite{b7}, \cite{b8}. 

In this study, we optimize a bank of IIR filters over the frequency regions in which most of the heave motion energy is expected to lie. The optimization is carried out using a large synthetically generated dataset that reflects realistic conditions by incorporating wave characteristics, vessel size effects, and vibration-induced as well as slowly varying inertial sensor errors. The contents of the following sections are organized as follows. In Section \ref{Sec2}, the procedure for generating the synthetic dataset is described in detail. Section \ref{Sec3} is dedicated to the optimization of the filter parameters, each of which is associated with a different group of heave signals categorized by their estimated mean peak periods. In Section \ref{Sec4}, the simulation results are presented, and Section \ref{Sec5} concludes the paper.

\section{Heave Dataset Generation Scheme}
\label{Sec2}
\begin{figure}[htbp]
	\centerline{\includegraphics[width=1\columnwidth]{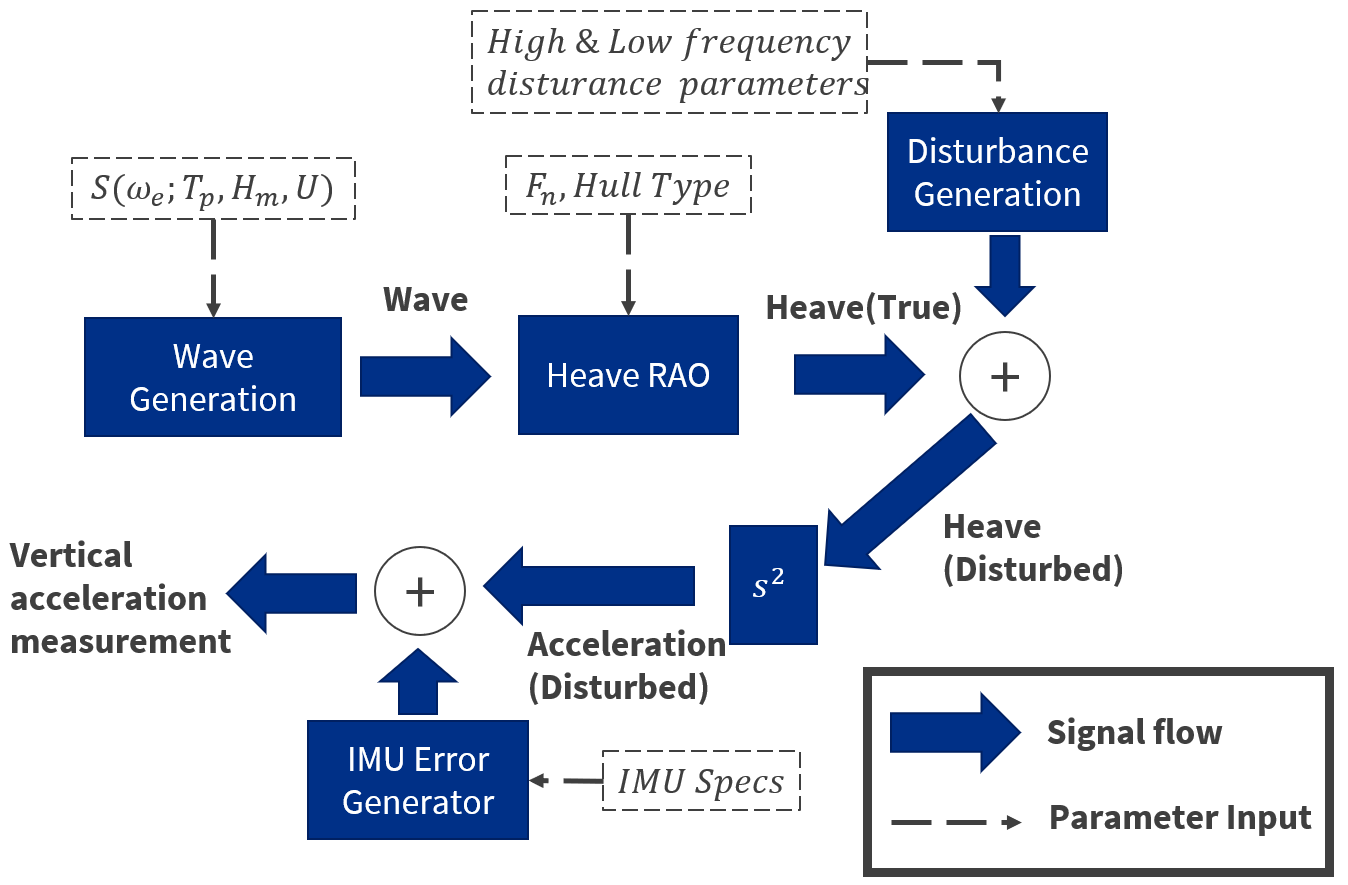}}
	\caption{Heave dataset generation framework.}
	\label{fig:HeaveDatasetGen}
\end{figure}

In this study, the term dataset refers to a set of tuples, each consisting of the vertical displacement (heave) of a ship and the corresponding acceleration measurements. Fig. \ref{fig:HeaveDatasetGen} illustrates the dataset generation framework, with each block discussed in the following subsections.

\subsection{Wave Generation}

Waves are formed by wind blowing over the sea surface, transferring energy from the atmosphere to the water. At sea, the height of a point above mean waterline level can be expressed as a sum of sinusoidal harmonics with different amplitudes($A_k$), frequencies($\omega_k$) and phases($\phi_k$) as given in \eqref{eq:waveSum} where $k$ denotes the harmonic number.

\begin{equation}
	h_w(t)=\sum_{k=0}^\infty A_k\sin(\omega_k t+\phi_k)
	\label{eq:waveSum}
\end{equation}
Since sea waves are irregular in nature, they are characterized by a wave energy spectrum (WES), denoted by  $S(\omega)$, which describes the distribution of wave energy across frequencies \cite{b16}. The expression in \eqref{eq:moment} defines the (n)-th spectral moment, where the zeroth, second, and fourth moments correspond to the variances of wave elevation, wave velocity, and wave acceleration, respectively.

\begin{equation}
	m_n = \int_{0}^{\infty} \omega^{n} S(\omega)\, d\omega
	\label{eq:moment}
\end{equation}

\begin{figure}[htbp]
	\centerline{\includegraphics[width=\columnwidth]{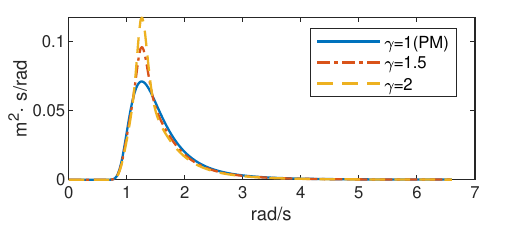}}
	\caption{JONSWAP WES's with different peakedness factors.}
	\label{fig:Jonswappeak}
\end{figure}

Among common WES models, Pierson–Moskowitz (PM) and Joint North Sea Wave Project (JONSWAP) are widely used for fully developed and fetch-limited sea states, respectively \cite{b9}, \cite{b10}. As reported in \cite{b11}, JONSWAP spectra with peakedness factors between 2 and 3 provide a good representation of wave conditions in the Mediterranean and Black Sea; therefore, this model is used in the present study. Figure~\ref{fig:Jonswappeak} illustrates JONSWAP spectra for different peakedness values. Wave characteristics are described by the mean peak period ($T_p$) and the significant wave height ($H_m$) which shape the corresponding WES.

\begin{figure}[htbp]
	\centerline{\includegraphics[width=\columnwidth]{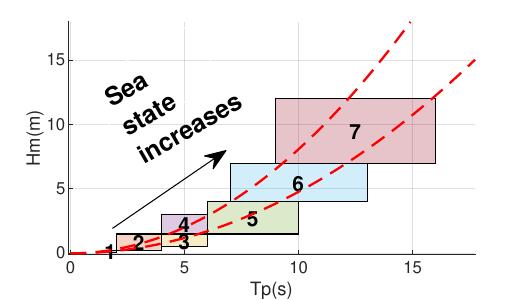}}
	\caption{Hm , Tp graph with wave parameter generation grids.}
	\label{fig:WaveGenReg}
\end{figure}

\begin{figure}[htbp]
	\centerline{\includegraphics[width=\columnwidth]{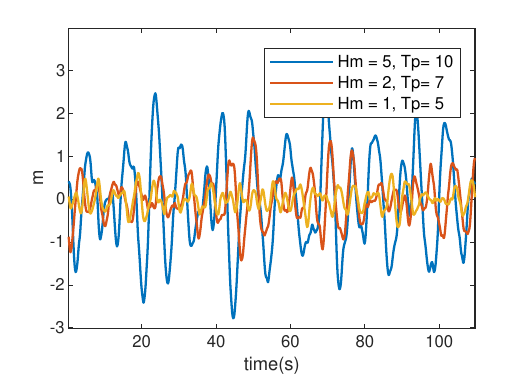}}
	\caption{Realization of waves in time domain with different wave parameters.}
	\label{fig:randomWaves}
\end{figure}

 There exists a positive correlation between $T_p$ and $H_m$. In \cite{b12}, the interval for $T_p$ is given as $\sqrt{6.5\times H^{(100)}} < T_p < \sqrt{11\times H^{(100)}}$, where $H^{(100)} = 1.9 H_m$. In Fig. \ref{fig:WaveGenReg}, the red dashed lines indicate these bounds for different $T_p$ values. Based on reported $H_m$ and $T_p$ values for the Mediterranean Sea in \cite{b13}, we construct an $H_m$–$T_p$ grid from which wave parameters are randomly sampled.  Random wave realizations are generated using the WAFO (Wave Analysis for Fatigue and Oceanography) MATLAB toolbox \cite{b18}. Fig. \ref{fig:randomWaves} shows examples of the generated wave signals.

Depending on the ship’s speed and the angle of the incoming wave $\beta$, the wave is perceived differently. Therefore, it is necessary to obtain WES with respect to encounter frequency $\omega_e$. In head seas condition ($\beta=\pi$) it is computed as given in \eqref{eq:SwE} \cite{b19}.
\begin{equation}
S\left(\omega_e\right)=\frac{S\left(\omega\right)}{\left(1+2\omega\frac{U}{G}\right)},\ \ \ \ \ for\ U,\omega\geq0
\label{eq:SwE}
\end{equation}
where $G$ denotes the gravitational acceleration and $U$ denotes the ship's  speed in knots.

\subsection{Heave RAO Generation}
The Heave Response Amplitude Operator (RAO) can be regarded as the ship’s transfer function, describing how incoming waves affect vertical motion. The heave RAO is commonly expressed as a function of the wavelength-to-ship-length ratio $\lambda/L$ and is a key outcome of hull design, as different hull types exhibit distinct responses for a given speed and wave encounter angle. In \cite{b14}, Ghassemi et al. derived heave RAOs for three hull types (Wigley, S60, and DDG) across various Froude numbers ($Fn$). The heave RAO can also be expressed as a function of the encounter frequency $\omega_e$ using \eqref{eq:omega_e}.
\begin{equation}
	\omega_e=\sqrt{\frac{2\pi G}{\lambda}}
	\label{eq:omega_e}
\end{equation}
where $G$ and $\lambda$ are the gravitational acceleration and wave length, respectively. The relationship between Froude number and wave length $\lambda$ is also expressed in \eqref{eq:fn}
\begin{equation}
	F_n=\frac{U}{\sqrt{GL}}
	\label{eq:fn}
\end{equation}
where $U$ is the ship speed in knots and $L$ is the ship length. To obtain the heave spectrum from the encountered WES and a heave RAO, one can use the formula given in \eqref{eq:Sheave}
\begin{equation}
	S_{heave}\left(\omega_e\right)=S(\omega_e)\left|H_{RAO}\left(\omega_e\right)\right|^2
	\label{eq:Sheave}
\end{equation}

\subsection{Disturbance and Noise Modeling}\label{sec:Disturbance}
The vertical acceleration is computed in a navigation computer as given in \eqref{eq:vertAccel}.
\begin{equation}
	a_z^N=\boldsymbol{e}_3^T\left(C_B^N\boldsymbol{f}^B-\left(2\boldsymbol{\omega}_{IE}^N+\boldsymbol{\omega}_{EN}^{\ N}\right)\times \boldsymbol{v}^N+\boldsymbol{g}^N\right)
	\label{eq:vertAccel}
\end{equation}
where $C_B^N$ denotes the direction cosine matrix transforming a vector resolution from the body frame B to the navigation frame N. The vectors $\omega_{IE}^N$   and $\omega_{EN}^N$  represent, respectively, the Earth’s rotation rate and the transport rate, both resolved in the N frame. The vector $v^N$ is the ship’s velocity, and $g^N$ is the plumb-bob gravity vector, with both expressed in the N frame. $f^B$ is the specific force vector measured by the accelerometer triad along its sensitive axes, expressed in the body frame. Lastly, $e_3^T$  denotes the row vector $ \left[0\ ,\ 0\ ,1\right].$ A navigation system computes the aforementioned vectors and matrix along with other quantities at high speed in order to update its kinematic states; position, velocity, and attitude. The computed $a_z^N$ contains not only the acceleration due to heave motion, but also high-frequency components caused by engine vibrations and structural resonances \cite{b15}, as well as slowly varying errors resulting from the combined effects of fixed bias, bias instability, sensor misalignment, and tilt errors in the vertical direction of the N frame. 

\begin{figure}[htbp]
	\centerline{\includegraphics[width=\columnwidth]{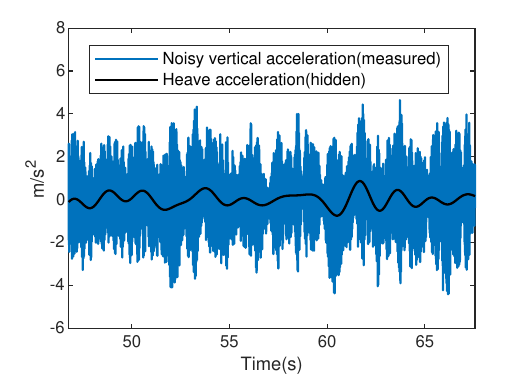}}
	\caption{Acceleration of the heave signal and its measured counterpart.}
	\label{fig:heaveAndVertaccel}
\end{figure}

To model these effects, IMU recordings collected during sea trials were analyzed in the frequency domain, where a low-frequency band of [0–0.05] Hz and a high-frequency band of [2–$f_{nq}$] Hz were selected, with $f_{nq}$ denoting the Nyquist frequency.  We utilized shaping filters to generate these high and low frequency noises and disturbances \cite{b17}. Figure \ref{fig:heaveAndVertaccel} illustrates an example of the generated heave acceleration and its measured counterpart containing high and low-frequency errors and disturbances.

\section{HEAVE FILTER OPTIMIZATION}
\label{Sec3}
The transfer function in \eqref{eq:GodhavnH} corresponds to the standard heave filter proposed by Godhavn \cite{b4}, which combines a high-pass filter with a double integrator to remove constant and low-frequency biases from the vertical acceleration measurements.
\begin{equation}
H\left(s\right)=\frac{p\left(s\right)}{a_z(s)}=\frac{s^2}{\left(s^2+2\zeta\omega_cs+\omega_c^2\right)^2}
\label{eq:GodhavnH}
\end{equation}
where $p(s)$ and $a_z(s)$ denote the vertical position and acceleration in the $s$-domain, respectively. $\zeta$ and $\omega_c$ are the damping ratio and cutoff frequency of the high-pass filter. A limitation of this standard heave filter is that it exhibits a large phase deviation from the ideal double-integrator response in the low-frequency region, which introduces delay in the estimated heave. To mitigate this phase issue, Richter \textit{et al}. propose a modified version of the standard filter by adding a pole–zero pair and a gain coefficient, as shown in \eqref{eq:Hsmod} \cite{b3}.
\begin{equation}
H\left(s\right)=\frac{p\left(s\right)}{a_z(s)}=\frac{s^2}{\left(s^2+2\zeta\omega_cs+\omega_c^2\right)^2}\frac{(s-s_z)}{(s-s_p)}K
\label{eq:Hsmod}
\end{equation}
The objective of the filter optimization is to determine a set of filter parameters $\omega_c$, $\zeta$, $p$, and $K$ that achieve the desired heave estimation performance. Rather than using a single parameter set for all conditions, the parameters are optimized separately for different heave classes, each corresponding to a distinct frequency band in which the dominant heave energy is expected to reside. These heave classes are defined based on the estimated peak period $T_p$ obtained via frequency-domain analysis, as described in the following subsection.

\subsection{Dominant Heave period estimation from vertical acceleration measurements}\label{Tpest}
Heave $T_p$ is estimated via discrete  Power Spectral Density (PSD) that is computed via Fast Fourier Transform (FFT) of a batch of vertical acceleration. FFT and PSD relationship is shown in \eqref{eq:Sk}
\begin{equation}
	S_{acc}\left[k\right]=\frac{X_{acc}\left[k\right]X_{acc}^\ast\left[k\right]}{N\ F_s},\ k\in[0,\ N-1]
	\label{eq:Sk}
\end{equation}
where $X_{acc}[k]$, $N$, and $F_s$ denote the FFT of the vertical acceleration at index $k$, the signal length in samples, and the sampling frequency, respectively. The frequency resolution $\Delta F$ is given in \eqref{eq:DeltaF}.
\begin{equation}
	\Delta F=\frac{F_s}{N}
	\label{eq:DeltaF}
\end{equation}
The relationships between acceleration PSD and velocity and displacement PSD are given in \eqref{eq:PSDvel}, and \eqref{eq:PSDpos}.
\begin{equation}
	S_{vel}\left[k\right]=\frac{S_{acc}\left[k\right]}{(2\pi \Delta F\cdot k)^2}
\label{eq:PSDvel}
\end{equation}
\begin{equation}
	S_{pos}\left[k\right]=\frac{S_{acc}\left[k\right]}{(2\pi \Delta F\cdot k)^4}
	\label{eq:PSDpos}
\end{equation}
As shown in \eqref{eq:PSDpos}, low-frequency components are amplified, leading to incorrect mean peak period estimation. This distortion in $S_{pos}[k]$ primarily results from spectral leakage in the FFT. To mitigate this, windowing is applied before FFT computation.

\begin{algorithm}[!t]
	\caption{Heave period estimation}
	\label{psCode:Tp}
	\begin{algorithmic}[1]
		\STATE \textbf{Input:} $a_z^N$ history,  $L$
		\STATE Compute $S_{\text{acc}}[k]$ (Eq.\eqref{eq:Sk})
		\STATE Compute $S_{\text{pos}}[k]$ (Eq.\eqref{eq:PSDpos})
		\STATE Compute the amplitude-weighted average frequency of the $L$ highest peaks in $S_{\text{pos}}[k]$.
		\STATE Take the inverse of the frequency value found in Step 4
		\STATE \textbf{Output:} $\widehat{T_p}$
	\end{algorithmic}
\end{algorithm}

\subsection{Optimization of filter parameters for different Tp intervals}
We optimized different heave filter parameters ($w_c$, $s_z$, $s_p$, $K$) for different $T_p$ intervals which are given in the following table.
\begin{table}[!t]
	\centering
	\caption{Category intervals of $T_p$}
	\begin{tabular}{c c c c}
		\hline
		\textbf{C1} & \textbf{C2} & \textbf{C3} & \textbf{C4} \\
		\hline
		[0--2.5s] & [2.5s--3.5s] & [3.5s--4.5s] & [4.5s--5.5s] \\
		\hline
		\textbf{C5} & \textbf{C6} & \textbf{C7} & \textbf{C8} \\
		\hline
		[5.5s--9s] & [9s--12s] & [12s--15s] & [15s--20s] \\
		\hline
	\end{tabular}
	\label{tab:TpCategories}
\end{table}
To optimize the filter coefficients, we first generate a large number of random heave signals and their corresponding vertical acceleration measurements following the procedure described in Section \ref{Sec2}. The generated dataset are then divided into training and test datasets. Each sample in the training set is categorized based on its estimated $T_p$ value according to Table \ref{tab:TpCategories}.
\begin{equation}
\begin{split}
	\mathcal{C}_{Training}^Z = \Big\{ &(\boldsymbol{a}_z^1,\, \boldsymbol{h}^1),\ (\boldsymbol{a}_z^2,\, \boldsymbol{h}^2),\ \ldots,\ (\boldsymbol{a}_z^M,\, \boldsymbol{h}^M) \Big\}, \\
	& Z \in \{1,2,3,\ldots,8\}
\end{split}
\label{eq:Ctrain}
\end{equation}
The expression in \eqref{eq:Ctrain} defines the set of acceleration–heave tuples contained in Category–Z, where $\boldsymbol{a}_z^i$ and $\boldsymbol{h}^i$ denote the i th realization of the vertical-acceleration measurement history and its corresponding true heave history, respectively, and M denotes the total number of members in the category where $i\in{1,2,3\ldots,M}$.
Let $\boldsymbol{\theta}^{(Z)}$ denote the filter parameter vector associated with category Z as given in \eqref{eq:ThetaZ}
\begin{equation}
\boldsymbol{\theta}^{(Z)}=\left[\begin{matrix}\omega_c&s_p&s_z&K\\\end{matrix}\right]^T
\label{eq:ThetaZ}
\end{equation}
In the remainder of the section, ($Z$) notation is omitted for clarity. Recall that the heave filter to be optimized is given in \eqref{eq:Hstheta}
\begin{equation}
	H\left(s;\boldsymbol{\theta}\right)=\frac{s^2}{\left(s^2+2\zeta\omega_cs+\omega_c^2\right)^2}\frac{(s-s_z)}{(s-s_p)}K
\label{eq:Hstheta}
\end{equation}
$H(s;\boldsymbol{\theta})$ is discretized using the zero-order hold method, yielding the discrete filter coefficients $\mathbf{B}$ and $\mathbf{A}$.
\begin{equation}
	D_{\Delta t}^{ZOH}\!\left\{H(s;\boldsymbol{\theta})\right\}
	= H[z;\boldsymbol{\theta}] = \frac{\sum_{i=0}^{q} B_i(\boldsymbol{\theta}) z^{-i}}{1 + \sum_{j=1}^{l} A_j(\boldsymbol{\theta}) z^{-j}}
\end{equation}
where $\Delta t$ is the sampling period of the signals in the dataset. The estimated heave is computed by filtering the  $a_z^N$  with the $H\left[z;\boldsymbol{\theta}\right]$ as specified in \eqref{eq:hout}.

\begin{equation}
	\hat{h}\left[n;\boldsymbol{\theta}\right]=\ -\sum_{j=1}^{l}{A_j\left(\boldsymbol{\theta}\right)\hat{h}\left[n-j;\boldsymbol{\theta}\right]}+\sum_{i=0}^{q}{B_i\left(\boldsymbol{\theta}\right)a_z\left[n-i\right]}
\label{eq:hout}
\end{equation}
The heave error for $\left(\boldsymbol{a}_z^i,\ \ \ \boldsymbol{h}^i\right)$ tuple is defined as:
\begin{equation}
	E^i\left[n\right]={\hat{h}}^i\left[n;\boldsymbol{\theta}\right]-h^i\left[n\right]\ ,\ for\ n\in[1,2,3\ldots,\ N]  
\end{equation}
where upper script i specifies the sample number in the category, and n denotes the discrete time index. The cost function to be minimized is defined in \eqref{eq:Jtot}.
\begin{equation}
	J_{tot}\left(\boldsymbol{\theta}\right)=\alpha J_{rms}\left(\boldsymbol{\theta}\right)+\left(1-\alpha\right)J_{peak}\left(\boldsymbol{\theta}\right),\ \ \ \alpha\in(0,1)
\label{eq:Jtot}
\end{equation}
where $J_{rms}(\boldsymbol{\theta})$ and $J_{peak}(\boldsymbol{\theta})$ are defined in \eqref{eq:Jrms} and \eqref{eq:Jpeak}, respectively, and M denotes the total number of acceleration-heave tuples within in the category dataset under consideration.

\begin{equation}
	J_{rms}\left(\boldsymbol{\theta}\right)=\frac{1}{M}\sum_{i=1}^{M}\sqrt{\frac{1}{N}\sum_{n=1}^{N}\left(E^i\left[n\right]\right)^2}
\label{eq:Jrms}
\end{equation}
\begin{equation}
	J_{peak}\left(\boldsymbol{\theta}\right)=\frac{1}{M}\sum_{i=1}^{M}\max{\left(|E^i\left[n\right]|\right)}
\label{eq:Jpeak}
\end{equation}
We employed simulated annealing, a stochastic, black-box optimization method, to minimize the total cost $J_{tot}\left(\boldsymbol{\theta}\right)$ defined in \eqref{eq:Jtot}  More than 3000 distinct heave–acceleration tuples were generated to optimize each filter corresponding to its respective heave category. In Table \ref{tab:theta_bounds}, the bounds of the filter parameters are provided. Since each heave category has its own parameter set, the optimization is performed independently for each category.
\begin{table}[!t]
	\centering
	\caption{Parameter bounds for $\theta$}
	\begin{tabular}{c c c c c}
		\hline
		& $\theta[1]:\omega_c$ & $\theta[2]:s_p$ & $\theta[3]:s_z$ & $\theta[4]:K$ \\
		\hline
		Upper Limit & 0.8   & -0.1 & -0.1 & 1   \\
		Lower Limit & 0.001 & -5   & -6   & 0.3 \\
		\hline
	\end{tabular}
	\label{tab:theta_bounds}
\end{table}
\section{Simulation Results}
\label{Sec4}
For evaluation, the proposed method was first tested on a test dataset consisting of different types of heave segments. Then, the algorithm was embedded in the navigation software and evaluated in real time on a 6-DoF motion simulator.
\begin{table}[t]
\centering
\caption{Heave Types}
\label{tab:heave_segments}
\begin{tabular}{lccccc}
\hline
 & Type 1 & Type 2 & Type 3 & Type 4 & Type 5 \\
\hline
$H_m$ [m] & 0.3& 0.8 & 1.2 & 1.7 & 2.0 \\
$T_p$ [s] &2.5 & 5.5 & 6.5 & 8.0 & 10.0  \\
\hline
\end{tabular}
\end{table}
\subsection{Offline Test with Synthetic Dataset}
\begin{figure}[htbp]
	\centerline{\includegraphics[width=\columnwidth]{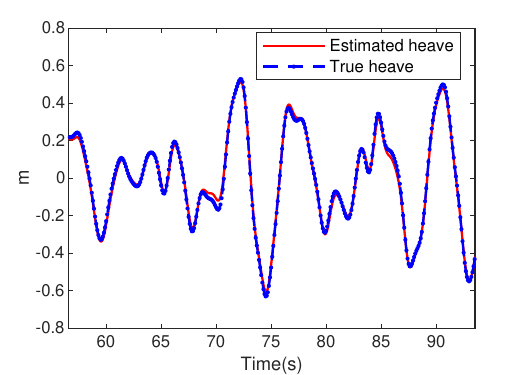}}
	\caption{True vs estimated heave.}
	\label{fig:Hvzoom}
\end{figure}

\begin{figure}[htbp]
	\centerline{\includegraphics[width=\columnwidth]{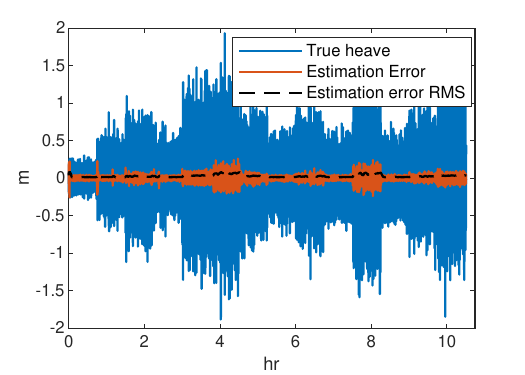}}
	\caption{True heave and heave estimation error.}
	\label{fig:HvvsHverr}
\end{figure}

In offline test, 10.5 hour long test dataset was formed by concatenating 45-minutes long heave segments whose parameters ($T_p$ and $H_m$) correspond to one of the heave types listed in Table \ref{tab:heave_segments}. All heave filters run in parallel, and the final heave output is determined by the estimated $T_p$. In Fig. \ref{fig:Hvzoom}, a small thirty-second part of the estimated heave and true heave is shown.   In Fig. \ref{fig:HvvsHverr} , the true heave is presented along with the estimation error and the associated 3-minute long running RMS for the entire test dataset.

\begin{figure}[htbp]
	\centerline{\includegraphics[width=\columnwidth]{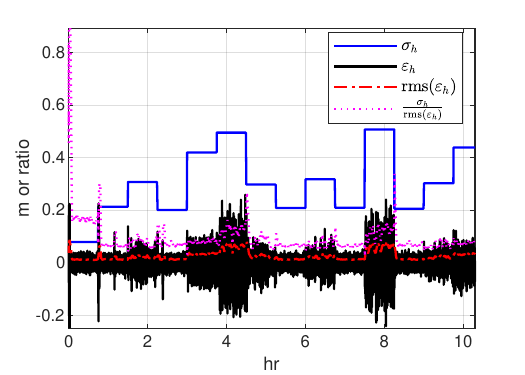}}
	\caption{Heave estimation error(black), True heave motion $1\sigma$(blue), Running RMSE(red), ratio of RMSE to heave $1\sigma$(purple).}
	\label{fig:Heaverms2errorrmsrat }
\end{figure}

\begin{figure}[htbp]
	\centerline{\includegraphics[width=\columnwidth]{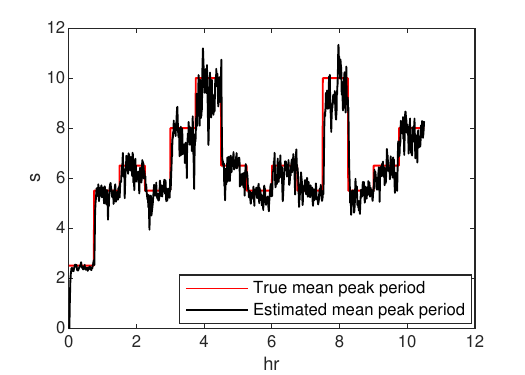}}
	\caption{True vs estimated mean peak heave period.}
	\label{fig:TpEstComp}
\end{figure}

In Fig. \Ref{fig:Heaverms2errorrmsrat }, $\sigma_h$  denotes standart deviation of heave motion that is fixed at each segment whereas $\epsilon_h$ denotes heave estimation error. Red dashed  line represents the 3-min running RMS of the estimation error. Purple dashed line is the ratio of $rms(\epsilon_h)$ to $\sigma_h$ with an average value of $10\%$. In other words, proposed framework, manages to yield RMSE that is one-tenth of the one standard deviation of the actual heave motion. This ratio corresponds to approximately $2.5\%$  when expressed as the ratio of $rms(\epsilon_h)$ to  $H_m$ since the significant heave height can be approximated by $4\times\sqrt{m_0}$ where $m_0$ is the variance of the heave motion \cite{b16}.  Mean peak period estimation graph is shown Fig. \ref{fig:TpEstComp} along with the true mean peak period of the generated dataset. The RMSE associated to $T_p$  estimation is $0.55$s.

\subsection{Real-Time Validation on a 6-DoF Motion Platform}
\begin{figure}[htbp]
	\centerline{\includegraphics[width=\columnwidth]{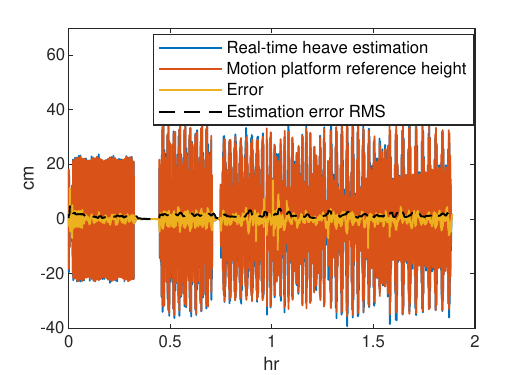}}
	\caption{Real-time test result.}
	\label{fig:Motionplatform}
\end{figure}

To validate the proposed algorithm, real-time tests were conducted using a 6-DoF motion platform. The platform was commanded to perform a 105-minute vertical motion profile mimicking heave motion, with the mean peak period varying from 5 s to 12 s. Fig. \ref{fig:Motionplatform} shows the motion platform vertical position and the estimated heave output of the navigation system, along with the estimation error and the running RMS. Given the physical displacement limits of ±40 cm, the standard deviation of the actual vertical motion over the entire simulation was 15 cm, whereas the heave RMSE was 1.5 cm, which is consistent with the results obtained from the offline test.

\section{Conclusion}
\label{Sec5}
In this study, we propose a heave motion estimation framework that employs a bank of IIR filters whose parameters are stochastically optimized using a realistically generated synthetic heave dataset composed of heave–acceleration tuples. Performance results from both online and offline tests demonstrate that the proposed approach adapts to changing wave conditions and achieve accurate heave estimation, with an RMSE on the order of one-tenth of the standard deviation of the true heave motion.

\end{document}